\documentclass[a4paper]{article}
\usepackage{amscd,amsmath,amssymb}
\usepackage{yhmath}

\newcommand{\cH}{{\cal H}}

\newcommand{\cQ}{{\cal Q}}
\newcommand{\cbQ}{\overline{\cal Q}}

\newcommand{\bQ}{{\overline Q}{}}
\newcommand{\bS}{{\overline S}{}}

\newcommand{\bI}{{\overline I}{}}

\newcommand{\bxi}{{\bar\xi}}

\newcommand{\bpsi}{{\bar\psi}{}}
\newcommand{\brho}{{\bar\rho}{}}

\newcommand{\cN}{{ {\cal N}   }}

\newcommand{\tPi}{{\widetilde \Pi}}

\newcommand{\sfrac}[2]{{\textstyle\frac{#1}{#2}}}

\def\im{{\rm i}}

\newcommand{\be}{\begin{equation}}
\newcommand{\ee}{\end{equation}}
\newcommand{\bea}{\begin{eqnarray}}
\newcommand{\eea}{\end{eqnarray}}
\newcommand{\ba}{\begin{array}} \newcommand{\ea}{\end{array}}

\newcommand{\und}{\qquad\textrm{and}\qquad}
\newcommand{\nn}{\nonumber}

\newcommand{\p}[1]{(\ref{#1})}

\begin{document}
\thispagestyle{empty}

\begin{flushright}
\end{flushright}
\vspace{2cm}

\begin{center}
{\huge\bf Extended supersymmetric Calogero model}
\end{center}
\vspace{1cm}

\begin{center}
{\LARGE\bf  Sergey Krivonos${}^{a}$, Olaf Lechtenfeld$^b$, \\[8pt]
Alexander Provorov${}^{a,c}$ and Anton Sutulin${}^a$}
\end{center}

\vspace{0.2cm}

\begin{center}
{${}^a$ \it
Bogoliubov  Laboratory of Theoretical Physics,\\
Joint Institute for Nuclear Research,
141980 Dubna, Russia}\vspace{0.3cm}

${}^b$ {\it
Institut f\"ur Theoretische Physik and Riemann Center for Geometry and Physics, \\
Leibniz Universit\"at Hannover,
Appelstrasse 2, 30167 Hannover, Germany}\vspace{0.3cm}

{${}^c$ \it Moscow Institute of Physics and Technology (State University), \\
Institutskii per. 9, Dolgoprudny, 141701 Russia}
\vspace{0.5cm}

{\tt krivonos@theor.jinr.ru, lechtenf@itp.uni-hannover.de,\\
aleksanderprovorov@gmail.com, sutulin@theor.jinr.ru}
\end{center}
\vspace{3cm}

\begin{abstract}\noindent
We present a surprising redefinition of matrix fermions which brings the supercharges
of the $\cN$-extended supersymmetric $A_{n-1}$ Calogero model introduced in \cite{KLS1}
to the standard form maximally cubic in the fermions. The complexity of the model is
transferred to a non-canonical and nonlinear conjugation property of the fermions.
Employing the new cubic supercharges, we apply a supersymmetric generalization
of a ``folding'' procedure for $A_{2n-1}\oplus A_1$ to explicitly construct
the supercharges and Hamiltonian for arbitrary even-$\cN$ supersymmetric extensions
of the $B_n$, $C_n$ and $D_n$ rational Calogero models. We demonstrate that all
considered models possess a dynamical $osp(\cN|2)$  superconformal symmetry.
\end{abstract}


\newpage
\setcounter{page}{1}
\setcounter{equation}{0}
\section{Introduction}
\setcounter{equation}{0}
The $A_{n-1}$ Calogero Hamiltonian~\cite{CM},
describing one-dimensional particles with inverse-square pairwise interactions,
\be\label{1}  \vphantom{\bigg|}\smash{
H= \frac{1}{2} \sum_i^n p_i^2 + \frac{1}{2} \sum_{i\neq j}^n \frac{g^2}{(x_i-x_j)^2}\ ,
}
\ee
plays a significant role  in mathematical and theoretical physics.
Being the prime example of an integrable and solvable many-body system,
it appears in many areas of modern mathematical physics,
from high-energy to condensed-matter physics
(see e.g.~the review~\cite{Poly1} and refs.~therein).
An intriguing hypothesis suggests that the large-$n$ limit of an $n$-particle
$\cN{=}4$ superconformal rational Calogero model provides a microscopic description of
the extremal Reissner-Nordstr\"{o}m black hole in the near-horizon limit~\cite{BH}.
Since then, the task of constructing an (at least) $\cN{=}4$ supersymmetric $n$-particle
rational Calogero model has been the subject of a number of papers \cite{scm1}--\cite{scm10},
however with only partial success. Despite the simplicity of the Hamiltonian \p{1},
all attempts to find an $\cN{=}4$ supersymmetric version beyond the four-particle case
were unsuccessful. In contrast, the $\cN{=}2$ supersymmetric Calogero model has been found
many years ago~\cite{FM,BTW}.

The first attempt to construct an $\cN{=}4$ supersymmetric extension was performed
by Wyllard~\cite{scm2} with a discouraging result. Indeed, it was argued that such a system
does not exist at all. The next important step was taken in~\cite{scm5,scm6} where
the supercharges and Hamiltonian were explicitly constructed for the $\cN{=}4$
supersymmetric three-particle Calogero model. It was also shown that Wyllard's obstruction
can be interpreted as a quantum correction, so in the classical limit the Hamiltonian \p{1}
could be obtained. Unfortunately, beyond three particles the component description
in the Hamiltonian formalism of~\cite{scm5,scm6} leads to a system of nonlinear equations
for which even a proof of existence of solutions is rather nontrivial. Specifically, enlarging the
conformal algebra to the supersymmetric case imposes severe constraints on the interactions,
which are not easily solved. Firstly, there is a nonzero prepotential $F$ which must obey
a system of quadratic homogeneous differential equation of third order known as the
Witten--Dijkgraaf--Verlinde--Verlinde (WDVV) equations~\cite{wdvv1,wdvv2}.
The general solution to the WDVV equations is unknown, but various classes based on
(deformed) Coxeter root systems have been found (see e.g.~\cite{wdvv3}). Secondly,
another prepotential $U$ is subject to a system of linear homogeneous differential equations
of second order in a given $F$ background. For known $F$~solutions and without adding
harmonic spin variables, a nonzero $U$ has been found for only up to three particles, 
where the WDVV equations are still empty. A detailed discussion of the supersymmetrization 
of Calogero models can be found in the review~\cite{FIL}.

A common property of all these models is the limited number
of fermionic components accompanying the bosonic coordinates $x_i$
(four fermions for each $x_i$ in the case of $\cN{=}4$).
A different approach to supersymmetric Calogero-like models has been proposed
in~\cite{scm8,nCal3,nCal2,nCal1}.
Starting from a supersymmetrization of the Hermitian matrix model,
the resulting matrix fermionic degrees of freedom are packaged in $\cN{=}4$ superfields.
A similar extended set of fermions appeared in~\cite{nCal4},
however their bosonic sector contains no interaction.
Inspired by these results we developed a supersymmetrization of the free Hermitian matrix model
and constructed an $\cN$-extended supersymmetric $su(n)$ spin-Calogero model~\cite{KLS1}.
By employing a generalized Hamiltonian reduction adopted to the supersymmetric case,
we derived a novel rational $n$-particle Calogero model with an arbitrary even number
of supersymmetries. Like in the models of~\cite{scm8,nCal3,nCal2,nCal1,nCal4},
it features $\cN n^2$ rather than $\cN n$ fermionic coordinates and increasingly high
fermionic powers in the supercharges and the Hamiltonian.

While quite satisfactory from a mathematical point of view,
our new $\cN$-extended supersymmetric Calogero models~\cite{KLS1}
look very complicated for possible applications. The reason for this is expressions
$\sqrt{g + \Pi_{ii}}$, where $g$ is a coupling constant and $\Pi_{ij}\sim(\textrm{fermions})^2$,
present in the supercharges and the Hamiltonian (see Section~2 for details).
Since the $\Pi_{ij}$ are nilpotent, the Taylor expansion of the square root eventually terminates,
but already in the two-particle case with $\cN{=}4$ supersymmetry we encounter
the lengthy expression
\be \smash{
\sqrt{g+\Pi_{11}}=\sqrt{g}\,
\bigl( 1 +\sfrac{1}{2 g} \Pi_{11}-\sfrac{1}{8 g^2} \Pi_{11}^2
+\sfrac{1}{16 g^3} \Pi_{11}^3-\sfrac{5}{128 g^4} \Pi_{11}^4 \bigr)\ .
}
\ee
For $n$ particles the series will end with a term proportional to $\left(\Pi_{ii}\right)^{\cN (n-1)}$,
generating higher-degree monomials in the fermions, both for the supercharges and for the Hamiltonian.

The novelty of the present paper is a non-trivial redefinition of the matrix fermions,
which brings the supercharges of $\cN$-extended supersymmetric Calogero models~\cite{KLS1}
to the standard form, maximally cubic in the fermions. It is presented in Section~3.
The complexity of the initial supercharges is shifted to a non-canonical and nonlinear
conjugation property of the redefined fermions.
The simple form of the supercharges admits a supersymmetric generalization of
a ``folding'' procedure~\cite{OlPe,Poly3},
which relates the $A_{2n-1}\oplus A_1$ Calogero model with the $B_n$, $C_n$ and $D_n$ ones.
In Section~4 we provide a supersymmetric extension of the $B_n$, $C_n$ and $D_n$
Calogero models with an arbitrary even number of supersymmetries.
Section 5 explicitly demonstrates that all considered models possess
dynamical $osp(\cN|2)$ superconformal symmetry.

\setcounter{equation}{0}
\section{Extended supersymmetric Calogero model}
The starting point of our previous construction of the $\cN$-extended supersymmetric
Calogero model~\cite{KLS1} has been a supersymmetric extension of the
Hermitian matrix model \cite{sCal, GH, Poly2}, which includes the following set of fields,
\begin{itemize}
\item $n$ bosonic coordinates $x_i$, which come from the diagonal elements
of the Hermitian matrix $X$, and the corresponding momenta $p_j$ for $i,j=1,\ldots,n$;
\item off-diagonal elements of the matrix $X$, encoded in the angular momenta
$(\ell_{ij})^\dagger = \ell_{ji}$ with $\sum_i \ell_{ii}=0$ and non-vanishing Poisson brackets
\be\label{PB0}
\left\{ x_i, p_j\right\} = \delta_{ij}, \quad
\left\{ \ell_{ij}, \ell_{km}\right\} = \im\, \left( \delta_{im} \ell_{kj} - \delta_{kj} \ell_{im} \right) ;
\ee
\item fermionic matrices containing $\cN\,n^2$ elements $\rho^a_{ij}, \brho_{ij\,a}$
for $a=1,\ldots,\cN/2$ with $(\rho^a_{ij})^\dagger=\brho_{ji\, a}$ and brackets\footnote{
We identify the fermions $\psi^a_i, \bpsi_{i\,a}$ of~\cite{KLS1} with the diagonal part
of the fermionic matrices, $\psi^a_i = \rho^a_{ii}$ and $\bpsi_{i\,a} = \brho_{ii\,a}$.}
\be\label{PB0f}
\left\{ \rho^a_{ij} , \brho_{km\, b}\right\} = -\im \, \delta^a_b \delta_{im}\delta_{jk}.
\ee
\end{itemize}
Using these ingredients we have constructed the supercharges
\be\label{Q0}
Q^a= \sum_{i=1}^n p_i \rho^a_{ii} +\im \sum_{i \neq j}^n \frac{\left( \ell_{ij}+\Pi_{ij}\right) \rho^a_{ji}}{x_i-x_j} \und
\bQ_a= \sum_{i=1}^n p_i \brho_{ii\,a} - \im \sum_{i \neq j}^n \frac{\brho_{ij\,a}\left( \ell_{ji}+\Pi_{ji}\right) }{x_i-x_j}
\ee
obeying the $\cN$-extended super-Poincar\'{e} algebra
\be\label{NSP}
\left\{ Q^a , \bQ_b \right\} = - 2 \im\, \delta^a_b\, H \und
\left\{ Q^a, Q^b \right\}=\left\{ \bQ_a, \bQ_b \right\}=0
\ee
with the Hamiltonian
\be\label{H0}
H= \frac{1}{2}\sum_{i=1}^n p_i^2 + \frac{1}{2}\sum_{i \neq j}^n
\frac{\left( \ell_{ij}+\Pi_{ij}\right)\left( \ell_{ji}+\Pi_{ji}\right) }{\left(x_i-x_j\right)^2}\ ,
\ee
modulo the first-class constraints
\be\label{scon1}
\chi_i:=\ell_{ii}+\Pi_{ii} \approx  0 \quad \forall\; i \ .
\ee
Here, the fermionic bilinear
\be\label{Pi0}
\Pi_{ij} = \sum_{a=1}^{\cN/2} \;\sum_{k=1}^n \left( \rho^a_{ik}\brho_{kj\,a}+\brho_{ik\,a}\rho^a_{kj}\right)
\ee
obeys $\sum_i\Pi_{ii} =0$ and
provides a realization of the $su(n)$ algebra
\be\label{sunPi0}
 \left\{ \Pi_{ij}, \Pi_{km} \right\}=\im \left( \delta_{im} \Pi_{kj}-\delta_{kj}\Pi_{im}\right)\ .
\ee

We have chosen the simplest realization of the $su(n)$ generators $\ell_{ij}$,
namely in terms of semi-dynamical polar variables $r_i$ and $\phi_i$ through~\cite{KLS1}
\be\label{realiz0}
\ell_{ij} = -r_i r_j\,\textrm{e}^{\im (\phi_i-\phi_j)}+\sfrac{1}{n} \delta_{ij} \sum_{k=1}^n r_k\ ,
\ee
and resolved the constraints \p{scon1} via
\be\label{fincon}
r_i^2 \approx g+\Pi_{ii} \und \phi_i\approx 0 \qquad\mathrm{for}\quad i=1,\ldots,n\ ,
\ee
where $g$ is an arbitrary real constant.
The supercharges $Q^a$ and $\bQ_a$ \p{Q0} and the Hamiltonian $H$ \p{H0} then read
\be\label{finQH}
\begin{aligned}
{\widehat Q}^a \ &=\ \sum_{i=1}^n p_i \rho^a_{ii}\ -\ \im \sum_{i \neq j}^n \frac{\left(\sqrt{g+\Pi_{ii}}\sqrt{g+\Pi_{jj}}-\Pi_{ij}\right) \rho^a_{ji}}{x_i-x_j}\ , \\
{\widehat \bQ}_a \ &=\ \sum_{i=1}^n p_i \brho_{ii\,a}\ +\ \im \sum_{i \neq j}^n \frac{\brho_{ij\,a}\left( \sqrt{g+\Pi_{ii}}\sqrt{g+\Pi_{jj}}-\Pi_{ji}\right) }{x_i-x_j}\ , \\
\widehat{H} \ &=\ \frac{1}{2}\sum_{i=1}^n p_i^2\ +\ \frac{1}{2}\sum_{i \neq j}^n \frac{\left( \sqrt{g+\Pi_{ii}}\sqrt{g+\Pi_{jj}}-\Pi_{ij}\right)
\left( \sqrt{g+\Pi_{ii}}\sqrt{g+\Pi_{jj}}-\Pi_{ji}\right) }{\left(x_i-x_j\right)^2}\ .
\end{aligned}
\ee
They still form an $\cN$-extended super-Poincar\'{e} algebra \p{NSP}, thus describing
an $\cN$-extended supersymmetric rational Calogero model of type $A_{n-1}\oplus A_1$.

\setcounter{equation}{0}
\section{A new fermionic basis}
In terms of the fermions $\rho^a_{ij}$ and $\brho_{ij\,a}$,
the supercharges  ${\widehat Q}^a , {\widehat \bQ}_a$  and
the Hamiltonian $\widehat{H}$ \p{finQH} have a rather complicated structure due to the presence
of the square roots $\sqrt{g+\Pi_{ii}}$. Their Taylor expansion in powers of the fermionic
bilinear $\Pi_{ii}$ results in a long nilpotent tail in the supercharges as well as in the Hamiltonian.
It seems that this complicated structure is an intrinsic feature of the extended
supersymmetric Calogero models. However, this is not so.
As we will demonstrate, it is possible to redefine the fermionic components
in such a way as to bring the supercharges and Hamiltonian to the standard form,
with the fermions appearing maximally cubicly in the supercharges and quartically in the Hamiltonian.
The price to pay for this simplicity is a more complicated conjugation rule for the fermions.

The advertized transformation defines new fermions $\xi^a_{ij}$ and $\bxi_{ij\,a}$ as follows,
\be\label{xi}
\xi^a_{ij} = \frac{1}{\sqrt{g+ \Pi_{ii}}}\, \rho^a_{ij}\, \sqrt{g+ \Pi_{jj}} \und
\bxi_{ij\, a} = \frac{1}{\sqrt{g+ \Pi_{ii}}}\, \brho_{ij\,a}\, \sqrt{g+ \Pi_{jj}}.
\ee
Since it is a similarity transformation with a diagonal matrix,
the corresponding fermionic bilinears read
\be\label{Pixi}
\Pi^{\xi}_{ij}
= \sum_{a=1}^{\cN/2} \sum_{k=1}^n \left( \xi^a_{ik}\bxi_{kj\,a}+\bxi_{ik\,a}\xi^a_{kj}\right)
= \frac{1}{\sqrt{g+ \Pi_{ii}}}\, \Pi_{ij}\, \sqrt{g+ \Pi_{jj}},
\ee
so that the diagonal terms remain unchanged,
\be\label{Piii}
\Pi^{\xi}_{ii} = \sum_{a=1}^{\cN/2} \;\sum_{k=1}^n \left( \xi^a_{ik}\bxi_{ki\,a}+\bxi_{ik\,a}\xi^a_{ki}\right)
= \sum_{a=1}^{\cN/2} \;\sum_{k=1}^n \left( \rho^a_{ik}\brho_{ki\,a}+\brho_{ik\,a}\rho^a_{ki}\right) = \Pi_{ii}.
\ee
It is rather easy to check that the $\Pi^{\xi}_{ij}$ still generate
the same $su(n)$ algebra \p{sunPi0} as the $\Pi_{ij}$,
\be\label{sunPixi}
 \bigl\{ \Pi^{\xi}_{ij}, \Pi^{\xi}_{km} \bigr\}=\im \bigl( \delta_{im} \Pi^{\xi}_{kj}-\delta_{kj}\Pi^{\xi}_{im}\bigr)\ .
\ee
Moreover, the new fermions $\xi^a_{ij}$ and $\bxi_{ij\,a}$ \p{xi} obey the same
Poisson brackets~\p{PB0f} as the old ones,
\be\label{PB2}
\left\{ \xi^a_{ij}, \bxi_{km\, b}\right\} = -\im \, \delta^a_b \delta_{im}\delta_{jk}.
\ee

The main clue is that the supercharges ${\widehat Q}^a$ and $ {\widehat \bQ}_a$,
when rewritten in terms of $\xi^a_{ij}$ and $\bxi_{ij\,a}$,
acquire the standard structure
\be\label{Q2}
 {\widehat Q}^a = \sum_{i=1}^n p_i \xi^a_{ii}\ -\ \im \sum_{i \neq j}^n \frac{\bigl(g + \Pi^{\xi}_{jj} - \Pi^{\xi}_{ij}\bigr)\, \xi^a_{ji}}{x_i-x_j} \und
{\widehat \bQ}_a =\sum_{i=1}^n p_i \bxi_{ii\,a}\ +\ \im \sum_{i \neq j}^n \frac{\bigl(g + \Pi^{\xi}_{ii} - \Pi^{\xi}_{ji}\bigr)\, \bxi_{ij\,a}}{x_i-x_j}\ .
\ee
The rewritten Hamiltonian $\widehat{H}$ \p{finQH} also contains maximally four-fermion terms,
\be\label{H2}
\begin{aligned}
\widehat{H} \ &=\ \frac{1}{2}\sum_{i=1}^n p_i^2\ +\ \frac{1}{2}\sum_{i \neq j}^n \frac{\bigl( g+\Pi^{\xi}_{jj}-\Pi^{\xi}_{ij}\bigr)\,
\bigl( g+\Pi^{\xi}_{ii}-\Pi^{\xi}_{ji}) }{(x_i-x_j\bigr)^2}\ .
\end{aligned}
\ee

The complicated structure of the $\cN$-extended Calogero models is now hidden
in the conjugation rules for the new fermions and, consequentially, for $\Pi^{\xi}_{ij}$.
The previous rules
\be\label{con1}
\left( \rho^a_{ij}\right)^\dagger =  \brho_{ji\,a} \und  \left( \Pi_{ij}\right)^\dagger = \Pi_{ji}
\ee
induce on $\xi^a_{ij}$, $\bxi_{ij\,a}$ and $\Pi^{\xi}_{ij}$ the conjugation rules
\be\label{conxi}
\left( \xi^a_{ij}\right)^\dagger = \frac{g+\Pi^{\xi}_{jj}}{g+\Pi^{\xi}_{ii}}\, \bxi_{ji\,a} \und
\bigl( \Pi^{\xi}_{ij}\bigr)^\dagger = \frac{g+\Pi^{\xi}_{jj}}{g+\Pi^{\xi}_{ii}}\, \Pi^{\xi}_{ji}.
\ee
With respect to these new conjugation rules the supercharges \p{Q2} are conjugated to each other,
and the Hamiltonian \p{H2} is a Hermitian one for an $\cal N$-extended supersymmetric
Calogero model with $A_{n-1}\oplus A_1$ symmetry.

\setcounter{equation}{0}
\section{$B_n, C_n$ and $D_n$ supersymmetric Calogero models}
With the simple form \p{Q2} and~\p{H2}  of the supercharges ${\widehat Q}^a$ and ${\widehat \bQ}_a$
and the Hamiltonian $\widehat{H}$ one may apply a supersymmetric generalization of the
``folding'' procedure~\cite{OlPe} discussed by Polychronakos in~\cite{Poly3},
which relates the $A_{2n-1}\oplus A_1$ Calogero model with the $B_n$, $C_n$ and $D_n$ ones.

In the bosonic case this reduction\footnote{
We restrict to an even number $2n$ of particles because one additional particle
results only in the change of a coupling constant.}
imposes the following identification of the $2n$ coordinates $(x_i,x_{n+i})$ and
momenta $(p_i,p_{n+i})$ for $i=1,\ldots,n$~\cite{Poly3}:
\be\label{red1}
x_{2n+1-i} = -x_i \und p_{2n+1-i} = -p_i \qquad\textrm{for}\quad i \leq n.
\ee
The fermionic equations of motion suggest that \p{red1} should be supplemented
by the fermionic identifications
\bea\label{red2}
&&  \xi^a_{2n+1-i,j} = -\xi^a_{i,2n+1-j} = \xi^a_{2n+1-i, 2n+1-j} =-\xi^a_{i,j} \und \nn \\[4pt]
&& -\bxi_{2n+1-i,j\, a} = \bxi_{i,2n+1-j\, a} = \bxi_{2n+1-i, 2n+1-j\,a} =-\bxi_{i,j\,a}
\qquad\textrm{for}\quad i,j \leq n.
\eea
These relations are better visualized in matrix form,
\be
\begin{pmatrix}
(\xi^a_{i,j}) & (\xi^a_{i,n+j}) \\[4pt] (\xi^a_{n+i,j}) & (\xi^a_{n+i,n+j})
\end{pmatrix} = \begin{pmatrix}
\xi^a & \xi^a \cdot {\cal A} \\[4pt] -{\cal A}\cdot  \xi^a  & -{\cal A} \cdot \xi^a \cdot {\cal A}
\end{pmatrix} \quad\textrm{and}\quad
\begin{pmatrix}
(\bxi_{i,j\,a}) & (\bxi_{i,n+j\,a}) \\[4pt] (\bxi_{n+i,j\,a}) & (\xi_{n+i,n+j\,a})
\end{pmatrix} = \begin{pmatrix}
\bxi_a & -\bxi_a \cdot {\cal A} \\[4pt] {\cal A}\cdot  \bxi_a  & -{\cal A} \cdot \bxi_a \cdot {\cal A}
\end{pmatrix}
\ee
with the obvious notation for $n\times n$ matrices $\xi^a=(\xi^a_{i,j})$ and their conjugates
and an anti-diagonal $n\times n$ matrix
\be
{\cal A} = \left( \begin{smallmatrix}
0 & 0 & \cdots & 0 & 1 \\[4pt]
0 & 0 & \cdots & 1 & 0 \\
\vdots & \vdots & \adots & \vdots &\vdots \\[4pt]
0 & 1 & \cdots & 0 & 0 \\[4pt]
1 & 0 & \cdots & 0 & 0
\end{smallmatrix}
\right)\ .
\ee

As a consequence, the fermionic bilinears
\be
\widehat{\Pi}^\xi_{i,j} = \sum_{a=1}^{\cN/2} \;\sum_{k=1}^{n} \bigl(
\xi^a_{i,k}\bxi_{k,j\,a}+\bxi_{i,k\,a}\xi^a_{k,j} +\xi^a_{i,n+k}\bxi_{n+k,j\,a}+\bxi_{i,n+k\,a}\xi^a_{n+k,j}
\bigr)
\ee
and the analogous $\widehat{\Pi}^\xi_{n+i,j}$, $\widehat{\Pi}^\xi_{i,n+j}$
and $\widehat{\Pi}^\xi_{n+i,n+j}$ reduce as follows,
\be
\begin{pmatrix}
(\widehat{\Pi}^\xi_{i,j}) & (\widehat{\Pi}^\xi_{i,n+j}) \\[4pt]
(\widehat{\Pi}^\xi_{n+i,j}) & (\widehat{\Pi}^\xi_{n+i,n+j})
\end{pmatrix} = \begin{pmatrix}
2\,\Pi^\xi & -2\,\tPi^\xi\cdot{\cal A} \\[4pt] -2\,{\cal A}\cdot\tPi^\xi & 2\,{\cal A}\cdot\Pi^\xi\cdot{\cal A}
\end{pmatrix} \ ,
\ee
where $\Pi^\xi=(\Pi^\xi_{i,j})$ has been defined in \p{xi},
and the new $n\times n$ matrix $\tPi^\xi=(\tPi^\xi_{i,j})$ has the components
\be\label{tPi}
\tPi_{ij}^\xi=\sum_{a=1}^{\cN/2} \;\sum_{k=1}^{n } \left( \xi^a_{ik}\bxi_{kj\,a}-\bxi_{ik\,a}\xi^a_{kj}\right).
\ee
One may check that $\Pi^\xi_{ij}$ and $\tPi^\xi_{ij}$ form an $s( u(n)\oplus u(n))$ algebra
(remember that $\sum_i \Pi^\xi_{ii} =0$),
\be\label{susus}
\bigl\{ \Pi^{\xi}_{ij}, \Pi^{\xi}_{km} \bigr\}=\bigl\{ \tPi^{\xi}_{ij}, \tPi^{\xi}_{km} \bigr\}=
\im \bigl( \delta_{im} \Pi^{\xi}_{kj}-\delta_{kj}\Pi^{\xi}_{im}\bigr) \und
\bigl\{ \Pi^{\xi}_{ij}, \tPi^{\xi}_{km} \bigr\}=\im \bigl( \delta_{im} \tPi^{\xi}_{kj}-\delta_{kj}\tPi^{\xi}_{im}\bigr)
\ .
\ee

Before substituting the reduction \p{red1} and~\p{red2} into the supercharges \p{Q2}
one has to take into account that
\begin{itemize}
\item due to the non-zero Poisson brackets between the constraints \p{red1},
the Poisson brackets $\left\{x_i, p_j\right\}$ will be changed to Dirac brackets
$\left\{x_i, p_j\right\}_D=\frac{1}{2}\delta_{ij}$, so a standard bracket requires rescaling the $p_i$;
\item the fermions likewise have to be rescaled to regain the standard brackets \p{PB2};
\item the supercharges ${\widehat Q}^a$ and ${\widehat \bQ}_a$ must be rescaled
to yield the standard kinetic terms for the bosonic coordinates;
\item one may introduce two independent coupling constants $g$ and $g'$ in the reduced supercharges.
\end{itemize}
After taking care of these subtleties, one finally arrives at the supercharges
\bea\label{S3}
\cQ^a &=& \sum_{i}^n p_i \xi^a_{ii} -\im\sum_{i\neq j}^n \frac{\left( \Pi^\xi_{jj} -\Pi^\xi_{ij}\right)\xi^a_{ji}}{x_i-x_j}  +
\im \sum_{i,j}^n \frac{\left( \Pi^\xi_{jj} +\tPi^\xi_{ij}\right)\xi^a_{ji}}{x_i+x_j} +
 \im\,\frac{g}{2} \sum_{i\neq j}^n \left(\frac{\xi^a_{ij}-\xi^a_{ji}}{ x_i-x_j}+\frac{\xi^a_{ij}+\xi^a_{ji}}{ x_i+x_j}\right)\nn \\
&&+\im \,g'\,\sum_i^n \frac{\xi^a_{ii}}{x_i} \qquad\und \nn \\
\cbQ_a &=& \sum_{i}^n p_i \bxi_{ii\,a} -\im\sum_{i\neq j}^n \frac{\left( \Pi^\xi_{jj} -\Pi^\xi_{ij}\right)\bxi_{ji\,a}}{x_i-x_j}  -
\im \sum_{i,j}^n \frac{\left( \Pi^\xi_{jj} +\tPi^\xi_{ij}\right)\bxi_{ji\,a}}{x_i+x_j} +
 \im\,\frac{g}{2} \sum_{i\neq j}^n \left(\frac{\bxi_{ij\,a}-\bxi_{ji\,a}}{ x_i-x_j}-\frac{\bxi_{ij\,a}+\bxi_{ji\,a}}{ x_i+x_j}\right)\nn \\
&&-\im \,g'\, \sum_i^n \frac{\bxi_{ii\,a}}{x_i}
\eea
which, together with the Hamiltonian
\bea\label{H3}
\cH &=& \frac{1}{2} \sum_i^n p_i^2+\frac{1}{2}\sum_{i \neq j}^n \left[ \frac{\left( g+\Pi^\xi_{jj}-\Pi^\xi_{ij}\right)\left( g+\Pi^\xi_{ii}-\Pi^\xi_{ji}\right)}{(x_i-x_j)^2}+
\frac{\left( g+\Pi^\xi_{jj}+\tPi^\xi_{ij}\right)\left( g+\Pi^\xi_{ii}+\tPi^\xi_{ji}\right)}{(x_i+x_j)^2}\right]\nn \\
&&+\frac{1}{8}\sum_{i}^n \frac{\left( 2 g'+\Pi^\xi_{ii}+\tPi^\xi_{ii}\right)\left( 2 g'+\Pi^\xi_{ii}+\tPi^\xi_{ii}\right)}{x_i^2}\ ,
\eea
generate an $\cN$-extended super-Poincar\'{e} algebra~\p{NSP}.
The bosonic part of this Hamiltonian has the standard form for the $B_n$, $C_n$ and $D_n$
Calogero models,
\be\label{h3bos}
\cH_{\textrm{bos}} =\frac{1}{2} \sum_i^n p_i^2+\frac{g^2}{2}\sum_{i \neq j}^n \left[ \frac{1}{(x_i-x_j)^2}+
\frac{1}{(x_i+x_j)^2}\right]+\frac{{g'}^2}{2}\sum_{i}^n \frac{1}{x_i^2}.
\ee

To check that the supercharges $\cQ^a$ and $\cbQ_a$ in~\p{S3} and the Hamiltonian $\cH$ in~\p{H3}
form the algebra~\p{NSP} it is convenient  to treat the bilinears  $\Pi^\xi_{ij}$ and $\tPi^\xi_{ij}$
as independent objects, subject to the brackets~\p{susus} and obeying the following brackets with
the new fermions $\xi^a_{ij}$ and $\bxi_{ij\,a}$,
\bea
&& \left\{ \Pi_{ij}, \xi^a_{km}\right\} = \im\, \delta_{im} \xi^a_{kj}-\im\, \delta_{jk} \xi^a_{im}\ , \qquad
\left\{ \Pi_{ij}, \bxi_{km\,a}\right\} = \im\, \delta_{im} \bxi_{kj\,a}-\im\, \delta_{jk} \bxi_{im\,a}\ ,  \nn \\
&& \left\{ \tPi_{ij}, \xi^a_{km}\right\} = -\im\, \delta_{im} \xi^a_{kj}-\im\, \delta_{jk} \xi^a_{im}\ , \qquad
\left\{ \tPi_{ij}, \bxi_{km\,a}\right\} = \im\, \delta_{im} \bxi_{kj\,a}+\im\, \delta_{jk} \bxi_{im\,a}\ .
\eea
As usual, all otherwise straightforward computations heavily rely on the identity
\be\label{impeq}
\frac{1}{(x_i-x_j)(x_i-x_k)}+\frac{1}{(x_j-x_i)(x_j-x_k)}+\frac{1}{(x_k-x_i)(x_k-x_j)} = 0
\qquad \textrm{for} \quad i\neq j\neq k .
\ee

\setcounter{equation}{0}
\section{Superconformal invariance}
The new fermionic basis we introduced is convenient to demonstrate that all
extended supersymmetric Calogero models discussed in this paper
possess dynamical superconformal symmetry. In this section we write
$Q^a$, $\bQ_a$ and $H$ for the supercharges \p{Q2} or \p{S3} and
the Hamiltonian \p{H2} or \p{H3}, depending on the case.
Starting from the conformal conserved current
\be\label{K}
K = \sfrac{1}{2} \sum_{i=1}^n x_i^2 - t \sum_{i=1}^n x_i p_i +t^2 H,
\ee
the other conserved currents can easily be obtained by successive commutators of $K$
with the supercharges and the Hamiltonian.
In this manner we may find a complete list of the conserved currents:
\bea\label{currents}
D =  -\sfrac{1}{2} \sum_{i=1}^n x_i p_i + t H\ ,\qquad
J^a{}_b= - \sum_{i, j}^n \xi^a_{ij}\bxi_{ji\, b}\ ,\qquad
I^{ab}=- \sum_{i,j}^n \xi^a_{ij} \xi^b_{ji},\; \bI_{ab} =  \sum_{i, j}^n \bxi_{ij\,a} \bxi_{ji\,b}, \nn \\
S^a = \sum_{i=1}^n x_i \xi^a_{ii} - t Q^a\ , \qquad
\bS_a = \sum_{i=1}^n x_i \bxi_{ii\,a} - t \bQ_a\ . \qquad\qquad\qquad\qquad{}
\eea
Together with the  supercharges $Q^a$ and $\bQ_a$, the Hamiltonian $H$
and the conformal boost current $K$, these current build an $osp(\cN|2)$ superalgebra,
\bea\label{ospN2}
&& \left\{H, K\right\} = 2 D, \quad \left\{H, D \right\} = H, \quad \left\{ K,D\right\} = -K, \nn \\
&& \left\{ J^a{}_b, J^c{}_d\right\} =\im \left( \delta_b^c J^a{}_d -\delta^a_d J^c{}_b\right), \; \left\{ J^a{}_b, I^{cd}\right\} = \im \left( \delta_b^c I^{ad}-\delta_b^d I^{ac}\right), \;
  \left\{ J^a{}_b, \bI_{cd}\right\} = -\im \left( \delta^a_c \bI_{bd}-\delta^a_d \bI_{bc}\right),\nn \\
&& \left\{I^{ab}, \bI_{cd}\right\} = \im \left( \delta^a_c J^b{}_d  -\delta^a_d J^b{}_c -\delta^b_c J^a{}_d +\delta^b_d J^a{}_c \right), \nn\\
&& \left\{ D, Q^a\right\} = -\sfrac{1}{2} Q^a,\; \left\{ D, \bQ_a\right\} = -\sfrac{1}{2} \bQ_a, \quad
\left\{ D, S^a\right\} = \sfrac{1}{2} S^a,\;\left\{ D, \bS_a\right\} = \sfrac{1}{2} \bS_a,\; \nn \\
&& \left\{H, S^a\right\} = - Q^a,\;\left\{H, \bS_a\right\} = - \bQ_a,\quad \left\{K, Q^a\right\} = S^a,\; \left\{K, \bQ_a\right\} = \bS_a, \nn \\
&& \left\{ J^a{}_b, Q^c\right\} = \im\, \delta_b^c\, Q^a,\;\left\{ J^a{}_b, S^c\right\} = \im\, \delta_b^c\, S^a,\;
\left\{ J^a{}_b, \bQ_c\right\} = -\im\, \delta^a_c\, \bQ_b,\; \left\{ J^a{}_b, \bS_c\right\} = -\im\, \delta^a_c\, \bS_b , \nn \\
&& \left\{ I^{ab}, \bQ_c\right\} = -\im \left( \delta^a_c Q^b - \delta^b_c Q^a\right), \; \left\{ I^{ab}, \bS_c\right\} = -\im \left( \delta^a_c S^b - \delta^b_c S^a\right), \nn\\
&& \left\{ \bI_{ab}, Q^c\right\} = \im \left( \delta_a^c \bQ_b - \delta_b^c \bQ_a\right), \;  \left\{ \bI_{ab}, S^c\right\} = \im \left( \delta_a^c \bS_b - \delta_b^c \bS_a\right),  \nn\\
&&  \left\{ Q^a, \bQ_b\right\} = - 2 \im \delta^a_b H, \; \left\{ S^a, \bS_b\right\} = - 2 \im \delta^a_b K, \nn \\
&& \left\{ Q^a, \bS_b\right\} =  2\, \im\, \delta^a_b\, D +J^a{}_b, \; \left\{ S^a, \bQ_b\right\} =  2\, \im\, \delta^a_b\, D -J^a{}_b , \nn \\
&&  \left\{ Q^a, S^b\right\} =  I^{ab}, \; \left\{ \bQ_a, \bS_b\right\} = - \bI_{ab} . \;
\eea
The $J^a{}_b$ form an $u(\cN/2)$ subalgebra,
which is enhanced to an $so(\cN)$ subalgebra by adding the $I^{ab}$ and $\bI_{ab}$.

\section{Conclusions}
We developed further the Hamiltonian description of the $\cN$-extended supersymmetric
$A_{n-1}\oplus A_1$ rational Calogero model introduced in~\cite{KLS1}.
The crucial new feature is a particular redefinition
of the fermionic matrix degrees of freedom $\rho^a_{ij}$ and $\brho_{ij\,a}$ accompanying
the $n$ bosonic coordinates of the rational $A_{n-1}\oplus A_1$ Calogero model.
In terms of the new fermions $\xi^a_{ij}$ and $\bxi_{ij\,a}$ the supercharges forming
an $\cN$-extended super-Poincar\'{e} algebra are at most cubic in the fermions,
i.e.~they acquire the standard structure common to almost all known supersymmetric
mechanics~\cite{FIL}. The complicated structure of the initial supercharges and Hamiltonian
got traded for a quite complicated conjugation rule, which is an almost symbolic price to pay
for the drastic simplification.

The simple form of the supercharges allowed for a supersymmetric variant of the ``folding''
relating the bosonic $A$-type with $B_n, C_n$ and $D_n$ Calogero models~\cite{OlPe,Poly3}.
Performing such a reduction, we managed to formulate the resulting Calogero
models with $\cN$-extended supersymmetry. Finally, we demonstrated that these rational
$A_n, B_n, C_n$ and $D_n$ supersymmetric Calogero models possess dynamical
$osp(\cN|2)$  superconformal symmetry.

Let us list possible further developments:
\begin{itemize}
\item Exceptional Lie algebras:
extension of our analysis to models associated with $G_2$, $F_4$ or $E_n$.
For the $G_2$ Calogero model only the three-particle case with $\cN{=}4$ supersymmetry
has been elaborated~\cite{scm7}. A multi-particle $\cN{=}8$ mechanics with $F(4)$
superconformal symmetry has recently been constructed~\cite{FI1}.
\item Trigonometric models:
extension to the Calogero--Sutherland inverse-sine-square model. It is yet unclear whether for this
model the ``folding'' reduction~\cite{OlPe,Poly3} can be made to work in the supersymmetric case.
\end{itemize}

\vspace{0.5cm}

\subsection*{\bf Acknowledgements}
The work of S.K.\ was partially supported by RFBR grant 18-52-05002 Arm-a,
the one of A.S.\ by RFBR grants 18-02-01046 and 18-52-05002 Arm-a.
This article is based upon work from COST Action MP1405 QSPACE,
supported by COST (European Cooperation in Science and Technology).

\end{document}